\documentclass[reprint,superscriptaddress,aps,longbibliography]{revtex4-1}
\usepackage{amsmath,amssymb,amsfonts,amsthm}
\usepackage{dcolumn}
\usepackage{bm}
\usepackage{enumerate}
\usepackage{hyperref}
\hypersetup{colorlinks,allcolors=blue}
\usepackage{siunitx}
\usepackage{graphicx}
\usepackage{gensymb}
\graphicspath{{figs/}}

\begin{document}
    
\title{Physics-Informed Deep Image Prior Reconstruction of In-Plane Magnetization from Scanning NV Magnetometry}

\author{Zander Scholl} \affiliation{Materials Science Division, Argonne National Laboratory, Lemont, Illinois 60439, USA}

\author{Justin Woods} \affiliation{Materials Science Division, Argonne National Laboratory, Lemont, Illinois 60439, USA}

\author{Charudatta Phatak} \affiliation{Materials Science Division, Argonne National Laboratory, Lemont, Illinois 60439, USA} \affiliation{Department of Materials Science and Engineering, Northwestern University, Evanston, Illinois 60208, USA}

\author{Hanu Arava} \affiliation{Materials Science Division, Argonne National Laboratory, Lemont, Illinois 60439, USA}

\date{\today}
\begin{abstract}

Reconstructing magnetization in nanoscale magnetic thin films is essential for developing next-generation memory, sensors, and various spintronic technologies. However, this remains challenging due to the ill-posed nature of the stray field inverse problem, i.e., there are infinitely many magnetization solutions to a given stray field distribution. Here, we demonstrate that a physics-informed deep image prior (DIP) framework, using a simple convolutional autoencoder conditionally achieves a reasonable qualitative and quantitative reconstruction of complex in-plane magnetization patterns from scanning NV magnetometry. We find that the orientation of user-defined masks implemented to restrict the reconstruction solution space dramatically affects convergence. The optimal alignment of the mask improves the reconstruction signal-to-noise ratio by up to $\SI{3}{\decibel}$, thereby also serving as a diagnostic tool. The DIP approach requires no pre-trained datasets and is considered computationally less intensive as compared to supervised learning approaches. We analyze both Landau and dipole domain structures in lithographically patterned Permalloy nanostructures by incorporating experimentally-guided spatial constraints. Complementary magnetic force microscopy measurements were carried out to support the Scanning NV measurements. 

\end{abstract}

\maketitle

Spintronics is a rapidly emerging field, providing a low-energy alternative to existing CMOS technologies~\cite{Wolf2001,zutic2004spintronics,Baltz2018}. In order to facilitate the growth of the field, as we continue to discover new materials and applications, advanced magnetic imaging techniques capable of resolving complex magnetic textures at the nanoscale are essential~\cite{zutic2004spintronics,chappert2007future,hoffmann2013opportunities}. Scanning nitrogen-vacancy (NV) magnetometry (Figure 1) has emerged as a powerful tool, offering spatial resolution down to 20 nm and a field sensitivity in the range of $\mu\text{T}/\sqrt{\text{Hz}}$, but can reach the pico-Tesla regime under specialized conditions~\cite{taylor2008high,balasubramanian2008nanoscale,maletinsky2012robust,rondin2012nanoscale,hong2013nanoscale,rondin2014magnetometry,tetienne2015nature,wolf2015subpicotesla,gross2017real,casola2018probing,glenn2018high,thiel2019probing}. However, reconstructing the underlying magnetization from stray field measurements remains a challenging inverse problem~\cite{blakely1996potential,hug1998quantitative}. The magnetostatic transformation linking stray fields to magnetization is fundamentally ill-posed, as multiple magnetization configurations can produce identical stray field patterns, $B_{NV}(\mathbf{k}) = A M(\mathbf{k})$~\cite{lima2009obtaining,raissi2019physics} (Figure 1b). Here, $B_{NV}(\mathbf{k})$ corresponds to magnetic field measured along the NV quantization axis, $A$ is the forward operator tensor, and $M$ is the magnetization. 

Recently, artificial neural network approaches have demonstrated considerable progress in reconstructing a range of magnetic textures, including spatially-varying and topological spin structures. In particular, untrained physically informed neural networks and deep image prior (DIP) methods, which learn directly on individual images using physics-based loss functions, have shown promise for magnetization reconstruction~\cite{Dubois2022,Broadway2025}. These approaches circumvent the need for large training datasets by employing the implicit regularization properties of neural network architectures, where the network structure itself acts as a prior for natural image statistics. While these advances have successfully demonstrated reconstruction of various magnetic configurations including topological spin textures, systematic application to fundamental domain patterns in widely-used spintronic materials requires careful validation and optimization for technologically relevant systems such as racetrack memory. 

Addressing the need for efficient and accurate reconstruction in nanoscale ferromagnetic structures~\cite{parkin2008magnetic,allwood2005magnetic}, we demonstrate that deep image prior convolutional autoencoders can enable resource-efficient reconstruction of in-plane magnetic domain patterns from scanning NV magnetometry. We focus on Landau and dipolar domain configurations in lithographically patterned Permalloy nanostructures. Permalloy (Ni$_{80}$Fe$_{20}$) serves as a prototypical soft magnetic material with applications in magnetic storage, sensors, and spintronic devices due to its low coercivity and well-characterized domain behavior~\cite{hubert1998magnetic,bozorth1951ferromagnetism,chikazumi2009physics, Arava2026}. To our knowledge, this work is the first to experimentally validate DIP-based reconstruction for these specific magnetic textures using scanning NV magnetometry. Our approach diverges from previous studies that employed U-Net architectures with skip connections~\cite{Dubois2022}. Instead, we use a classical encoder-decoder network that more closely follows the original DIP methodology~\cite{ulyanov2018deep}, demonstrating that simplified architectures may be sufficient in structures with well-informed physics constraints. The reconstructed domain structures are validated through independent magnetic force microscopy measurements, confirming the accuracy of the neural network outcomes.

\begin{figure*}[t]
    \centering
    \includegraphics[width=\linewidth]{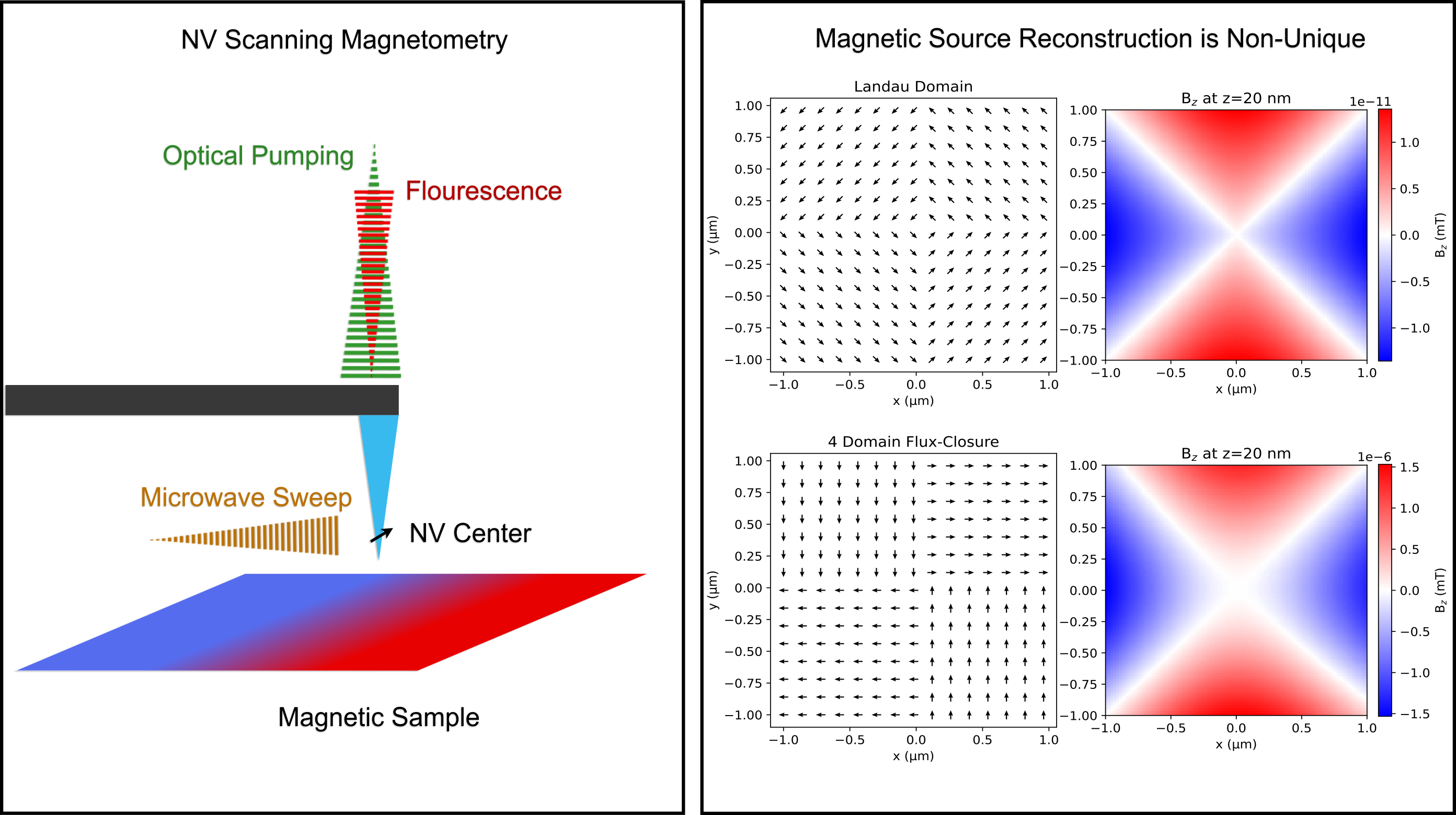}
    \caption{(A) Schematic of a scanning NV magnetometry experimental set up. The magnetic field strength is determined through optically-detected magnetic resonance (ODMR). The NV center is optically pumped with a green laser while a microwave antennae sweeps an applied RF field. (B) Simulated B field plots for two different magnetizations, a Landau domain and a 4 domain flux-closure, leading to nearly identical B-field projections. Illustrates the fundamental challenges associated with addressing an ill-posed problem.}
    \label{fig:enter-label}
\end{figure*}

\begin{figure}[h]
    \centering
    \includegraphics[width=\linewidth]{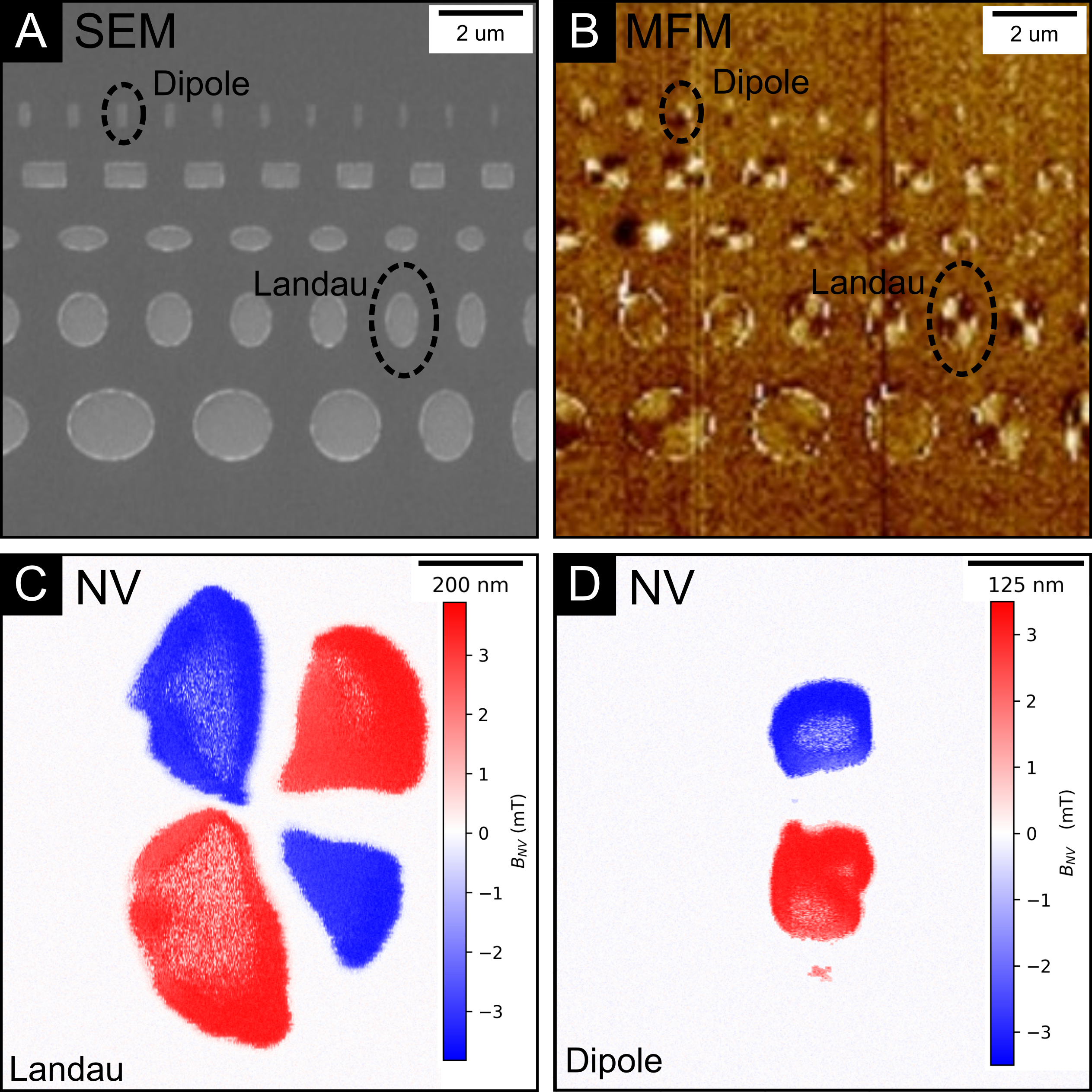}
    \caption{(A) SEM image of fabricated Permalloy nanostructures. (B) MFM images of selected ellipses, confirming the presence of Landau domain pattern (left) and dipole arrangement (right). (C-D) Scanning NV magnetometry B field maps of the same structures, here the NV axis is at 54.75 degrees relative to the sample normal, illustrating the four-lobed (Landau Domain) and two-lobed (Dipole Domain) stray field distributions.}
    \label{fig:sem_mfm_nv_maps}
\end{figure}

To demonstrate the capability of our reconstruction approach, we fabricated ellipse-shaped Permalloy nanostructures with varying aspect ratios (Figures 2A-B) using electron beam lithography (EBL). The nanostructures were deposited on a silicon substrate and had a thickness of approximately \SI{20}{nm}. In Figure~2A we show a representative scanning electron microscopy (SEM) image of the fabricated ellipses. Magnetic force microscopy (MFM) was employed to identify structures exhibiting distinct domain configurations. A total of 79 permalloy structures with varying aspect ratios were fabricated. We consider two nanostructures in the subsequent analysis, which were chosen based on high signal to noise ratio representing two prototypical domains observed in structures with in-plane magnetization: one exhibiting a Landau domain pattern and another with a uniform magnetization (dipole domain). The Landau domain ellipse measures \SI{1200}{nm} \(\times\) \SI{700}{nm}, while the dipole domain ellipse measures \SI{500}{nm} \(\times\) \SI{150}{nm}. The corresponding MFM images are presented in Figure~2B, which reveal the presence of domain walls and confirm the magnetic state of each structure. 

Quantitative stray field measurements were performed by scanning NV magnetometry using a commercial Quantum Scanning Microscope from QZabre LLC. A single NV center in a diamond probe was positioned above the sample with a standoff height of \SI{50}{nm}–\SI{70}{nm} to optimize signal-to-noise ratio which can degrade rapidly if the NV is exposed to large stray fields. The NV axis was oriented at \(54.75^\circ\) relative to the sample normal, enabling detection of both in-plane and out-of-plane magnetic field components. The measured \(B_{NV}\) field maps for both structures are presented in Figure~2C. In these maps, the Landau domain produces a characteristic four-lobed stray field distribution, while the dipole domain yields two distinct field maxima at the ends of the ellipses. The sources of the $B$ field tend to be regions where the magnetization is changing rapidly; typically these are domain walls. These features in the NV data are consistent with the domain configurations identified by MFM. These complementary imaging techniques provide direct experimental validation for the magnetic domain structures. Using $\textbf{mumax}^3$, we employed micromagnetic simulations to validate the domain structures (\cite{mumaxPaper}). Using material parameters for Permalloy and physical dimensions that matched the lithographically fabricated structures, we produced simulated magnetic domains with stray fields that qualitatively agreed with MFM and NV measurements. The spatial correspondence between MFM and scanning NV magnetometry confirms the reliability of the measured stray fields and the suitability of these structures for benchmarking our neural network-based reconstruction approach.


\begin{figure*}[t]
    \centering
    \includegraphics[width=\linewidth]{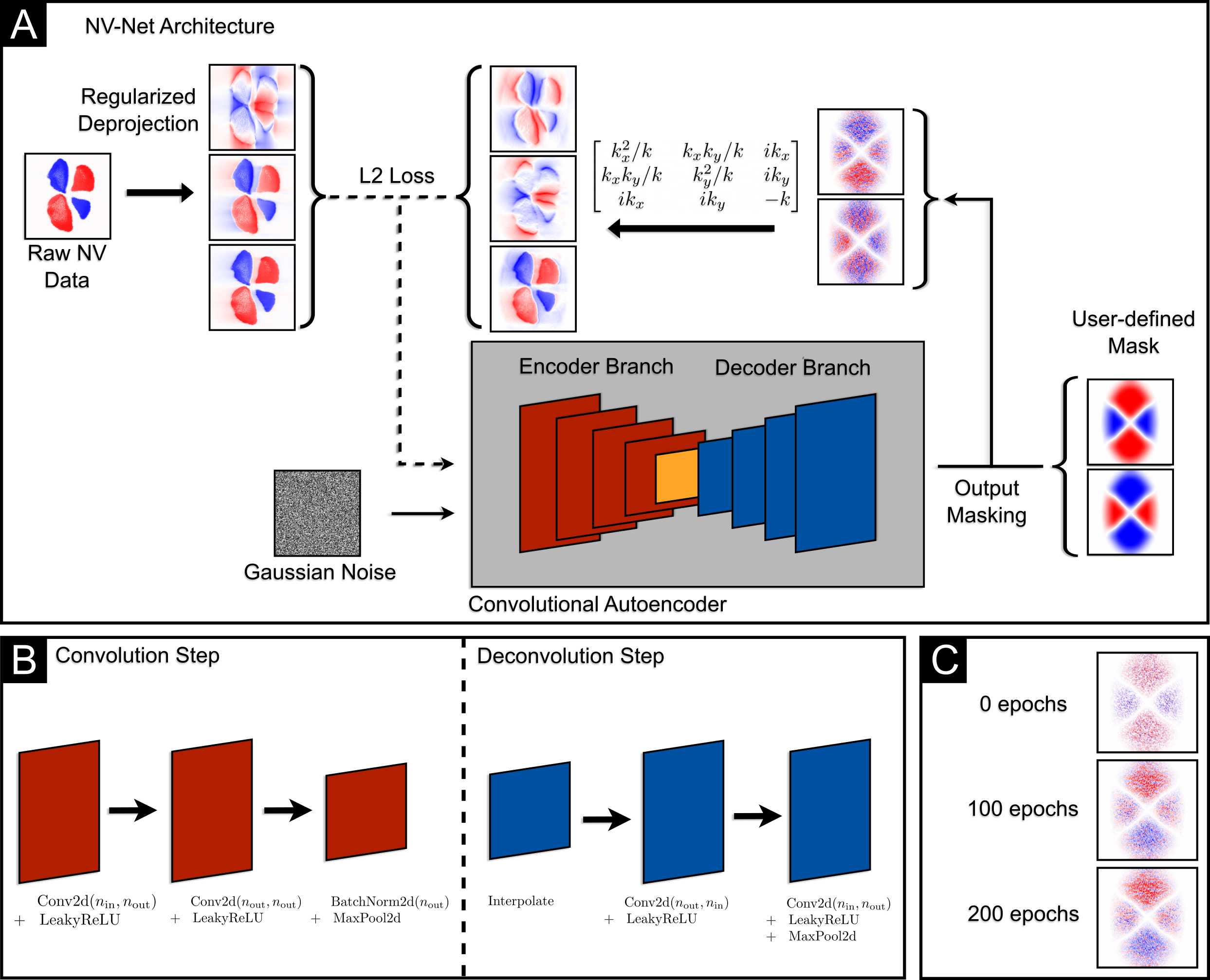}
    \caption{(A) Deep image prior reconstruction workflow showing the approach. Stray field data from scanning NV magnetometry is processed through a convolutional autoencoder that minimizes error in real-space between measured fields and forward-propagated outputs. (B) The encoder stage (teal) comprises paired convolutional layers (kernel size 5) with LeakyReLU activation, instance normalization, and max pooling for downsampling. Channel dimensions increase progressively through encoding stages. The decoder stage (yellow) uses identical convolutional parameters with bilinear upsampling, featuring separate channels for each magnetization component (M$_x$, M$_y$). We do not show M$_z$ because the network only considered in-plane domains. For each convolution and deconvolution step, a kernel size of 3 and a stride of 1 was used. The overall model utilized 2 convolution and deconvolution steps. (C) Reconstructed image of $M_x$ for 0 epochs, 100 epochs, and 200 epochs. The model used a domain mask rotation angle of $0^\circ$ and a standoff of $\SI{50}{\nano\meter}$.}
    \label{fig:enter-label}
\end{figure*}

\begin{figure*}[t]
    \centering
    \includegraphics[width=\linewidth]{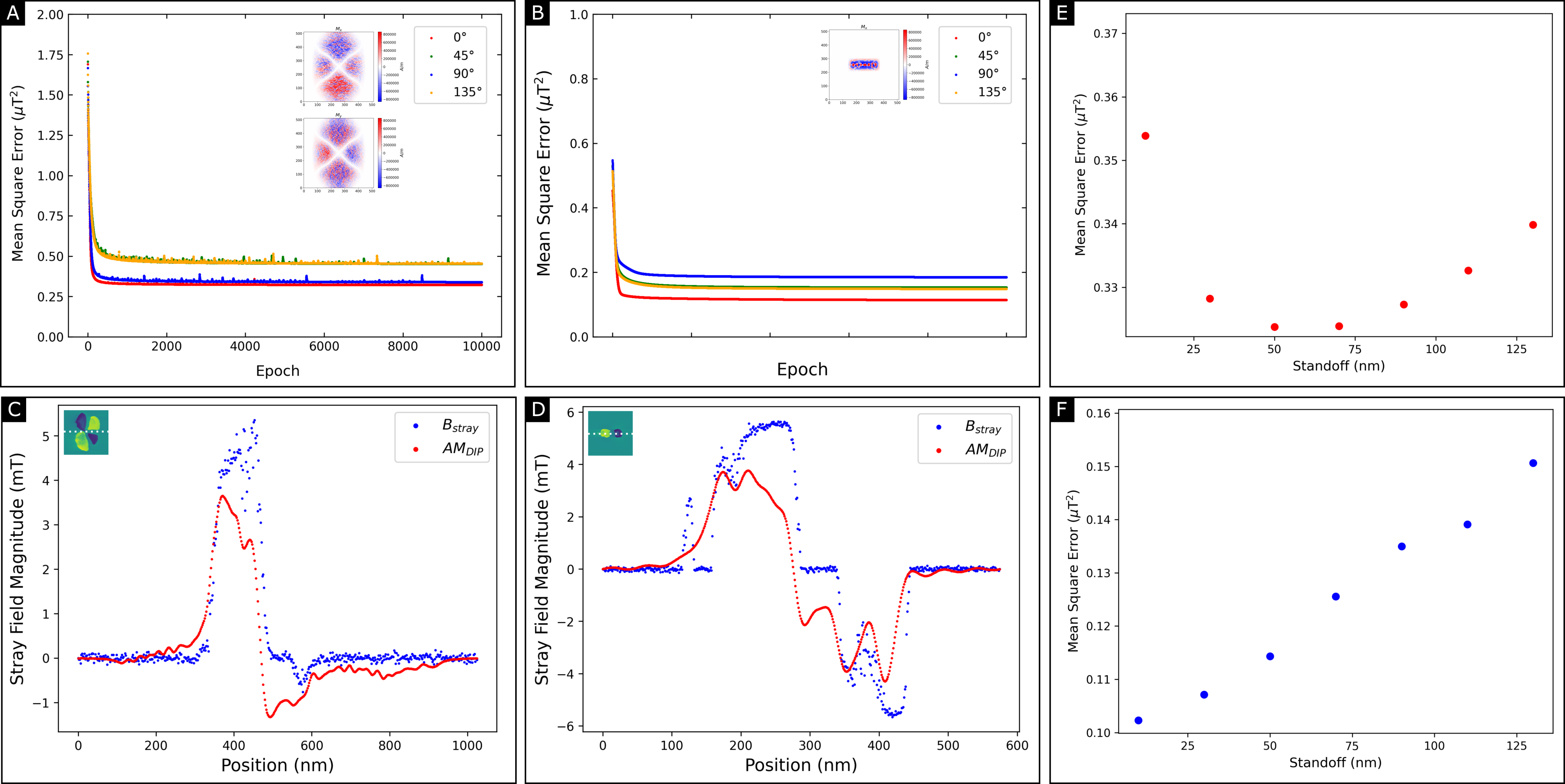}
    \caption{\textbf{ (A, B)} MSE as a function of training epochs for the (A) Landau and (B) Dipole domain reconstructions. Each curve represents a different orientation of the initialization mask, demonstrating that convergence is fastest for the correct orientation. In this figure, MSE is reported given that it was the actual cost function minimized during training whereas RMSE can be directly compared to the NV signal magnitude. \textbf{(C, D)} Line profiles comparing the measured stray field (Raw $B_{\text{NV}}$) with the final reconstructed field from the model, showing excellent agreement for both magnetic configurations. \textbf{(E, F)} MSE error as a function of standoff distance for Landau and Dipole domains, indicating the sensitivity of convergence as a function of tip to sample distance.}
    \label{fig:results_summary}
\end{figure*}

We reconstruct the sample magnetization using a deep image prior (DIP) approach implemented with a convolutional autoencoder (CAE) [Figure 3]. Unlike supervised learning methods that require extensive training datasets, DIP leverages the implicit regularization of a neural network's architecture to solve inverse problems on a single-instance basis~\cite{ulyanov2018deep,dong2019denoising}. The network receives a fixed random tensor (Gaussian noise, Figure 3A) as input and iteratively adapts its weights to generate a magnetization map that minimizes a physics-informed loss function (mean square error). This methodology could be effective for ill-posed inverse problems, such as reconstructing magnetization from stray fields, where multiple source distributions can produce identical field patterns. The physics-informed constraints are derived from the magnetostatic forward model connecting a sample's magnetization, $\vec{M}$, to the stray magnetic field, $\vec{B}$. In Fourier space, this relationship is a linear transformation given by~\cite{blakely1996potential,lima2009obtaining}:

\begin{equation}
\begin{pmatrix}
\tilde{B}_x \\
\tilde{B}_y \\
\tilde{B}_z
\end{pmatrix} = \frac{\mu_0}{2} e^{-kz}
\begin{pmatrix}
k_x^2/k & k_x k_y/k & i k_x \\
k_x k_y/k & k_y^2/k & i k_y \\
i k_x & i k_y & -k
\end{pmatrix}
\begin{pmatrix}
\tilde{M}_x \\
\tilde{M}_y \\
\tilde{M}_z
\end{pmatrix}
\label{eq:fwd_model_matrix}
\end{equation}

The NV sensor measures the projection of the magnetic field along its axis:
\begin{equation}
B_{NV} = \mathbf{B} \cdot \hat{\mathbf{n}}_{NV} = B_x n_x + B_y n_y + B_z n_z
\end{equation}
where $\hat{\mathbf{n}}_{NV} = (n_x, n_y, n_z)$ is the unit vector along the NV quantization axis. For our experimental configuration with the NV axis at 54.75° from the sample normal and aligned in the $x$-$z$ plane, this reduces to:
\begin{equation}
B_{NV} = B_x \sin(54.75^\circ) + B_z \cos(54.75^\circ)
\end{equation} To recover the magnetic stray field, the NV signal is deprojected using a 2D Fourier transform \cite{lima2009obtaining}:\begin{align}
 \mathcal{B}_x &= \frac{i k_x}{ik_xn_x + ik_yn_y+kn_z}\mathcal{B}_\textnormal{NV}, \\ 
 \mathcal{B}_y &= \frac{i k_y}{ik_xn_x + ik_yn_y+kn_z}\mathcal{B}_\textnormal{NV}, \\ 
 \mathcal{B}_z &= \frac{k}{ik_xn_x + ik_yn_y+kn_z}\mathcal{B}_\textnormal{NV}.
\end{align} Furthermore, the magnitude of the denominator in the above terms is regularized in order to prevent divergence at $k=0$. The optimal regularization parameter is iteratively determined by minimizing the mean square error between the reprojection of the computed stray field onto the NV axis and the original NV signal. Additionally, the network's objective is determine the magnetization ($\tilde{M}$) that minimizes the mean square error (MSE) between the Fourier transform of the measured field ($\mathbf{\tilde{B}}_{\text{NV,meas}}$) and the field predicted by the forward model ($\mathbf{\tilde{B}}_{\text{NV,model}}$):

\begin{equation}
\textnormal{MSE} = \sum_{\mathbf{r}} |\mathbf{B}_{\text{NV,meas}}(\mathbf{r}) - \mathbf{B}_{\text{NV,model}}(\mathbf{r})|^2
\label{eq:mse_error}
\end{equation}
where the sum extends over all pixels $\mathbf{r}$ in the real-space image. The reconstruction quality is additionally characterized by the signal-to-noise ratio (SNR):
\begin{equation}
    \textnormal{SNR}=10\log_{10}\Big(\frac{\textnormal{mean}(|\mathbf{A}\mathbf{M}|)}{\textnormal{MSE}}\Big),
\end{equation} where $\textnormal{mean}(|\mathbf{A}\mathbf{M}|)$ is the average magnitude of the forward-propagated stray field and $\textnormal{MSE}$ is defined as per Eq~\ref{eq:mse_error}. 

As stated previously, the magnetostatic inverse problem suffers from non-uniqueness. Loss minimization alone cannot resolve this ambiguity, necessitating additional constraints on model outputs. We address this challenge through two complementary mechanisms:

\textbf{Spatial Masking}: A user-defined tensor with elements ranging from $0$ to $1$ controls the spatial extent of allowed magnetization reconstruction through element-wise multiplication. This constraint prevents reconstruction in regions where no magnetic material exists. 

\textbf{Solution Initialization}: An assumed magnetization distribution guides local convergence, with convergence quality serving as a metric for initial guess accuracy. This approach is particularly critical for complex in-plane magnetizations, where domain patterns must be approximately known.

Simple collinear textures typically require only spatial masking, while complex in-plane magnetic textures necessitate both constraint types for successful reconstruction. 

To implement this, we use a CAE with two downsampling and two upsampling blocks, intentionally omitting the skip connections found in U-Net variants. This simplified structure provides a stronger implicit regularization, which is crucial for constraining the ill-posed problem and preventing overfitting to measurement noise~\cite{ulyanov2018deep}. The encoder blocks comprise of paired convolutional layers ($5 \times 5$ kernels, LeakyReLU activation), instance normalization, and max pooling. The decoder mirrors this using bilinear upsampling and provides separate output channels for each magnetization component ($M_x$, $M_y$, $M_z$). This depth was selected based on preliminary experiments showing diminishing returns and increased training instability in deeper networks. 
Training was performed using an AdamW optimizer with an initial learning rate of $7.5 \times 10^{-4}$ for the Landau domain and and $1 \times 10^{-4}$ for the dipole domain with a `ReduceLROnPlateau' scheduler (patience: 150 epochs, factor: 0.75, threshold: $1 \times 10^{-9}$). These scheduler parameters were chosen to facilitate thorough exploration of the loss landscape while avoiding premature convergence. To ensure reproducibility, all network training sessions were initiated from an identical Gaussian noise tensor initially generated with a random seed. 

To address the non-uniqueness of the solution, we employ a multi-stage output process to enforce physical constraints. The raw CAE outputs pass through a hyperbolic tangent activation layer (\texttt{Tanh}) and are then multiplied element-wise by a user-defined domain mask tensor spanning [-1, 1]. This step encodes prior knowledge of the magnetic texture and enforces zero magnetization outside the sample area. For complex in-plane patterns, we use masking informed by micromagnetic simulations to reflect features such as the presence of landau domains in permalloy structures of particular aspect ratios. The final output is then restricted in magnitude to the saturation magnetization to yield a quantitative result in physical units. We should note that without the physics informed masking, the network converges rapidly to the forward-propagated $B$ with very low error, but the reconstructed $M$ was rarely physically accurate. Therefore, in using DIP, physics informed constraints are a necessity. 

Furthermore, the choice of activation function for the final output layer had a non-trivial impact on model performance. Utilizing a \texttt{Sigmoid} activation function which bounds the model output to $[0, 1]$ forces agreement with the sign of the domain mask tensor at the cost of a decreased signal-to-noise ratio (SNR) and overall poorer convergence. In contrast, the \texttt{Tanh} activation function yielded superior convergence characteristics while allowing a qualitatively noisier magnetization solution. Given that the model output is highly sensitive to the domain mask (thus allowing the model to determine the relative likelihood of a given domain mask being quantitatively correct) we elected to use the \texttt{Tanh} activation function as the model solution must already be interpreted within the context of the domain mask.

To demonstrate our method's capabilities, we applied it to two challenging test cases in Permalloy thin films: a Landau domain pattern and a magnetic dipole configuration. To test the sensitivity of the model on the user-defined mask, we analyzed the model performance for several mask orientations. For the Landau, the angle of the magnetic domain boundaries was varied (Figure 4A) whereas the permitted magnetization region itself was rotated for the dipole (Figure 4B). The Landau domain reconstruction demonstrated better performance. With the optimal mask alignment (0° rotation), the model converged to an overall root mean square error of $\SI{0.56}{\milli\tesla}$  with a corresponding signal-to-noise ratio of $\SI{4.94}{\decibel}$ (Figure 4a). In contrast, the $90^\circ$ orientation converged to an overall RMSE value of $\SI{0.58}{\milli\tesla}$ while all other orientations converged above $\SI{0.67}{\milli\tesla}$.

In contrast to the model performance for experimental data, using the stray field and magnetization directly obtained from micromagnetic simulations of a Landau domain resulted in a substantially higher SNR of $\SI{22.33}{\decibel}$ with a final RMSE of $\SI{0.84}{\milli\tesla}$. Thus, model performance is highly dependent on prior assumptions which demonstrates the utility of error metrics such as SNR and RMSE as an assessment for the quality of predicted characteristics. 

For the magnetic dipole (Figure 4B), the method achieved an RMSE of $\SI{0.34}{\milli\tesla}$ when the mask orientation matched the true domain structure (0° rotation). The sensitivity suggests that mask orientation can serve as a powerful diagnostic tool, where the orientation yielding the fastest convergence indicates the underlying magnetic domain structure. Line profile analysis confirms reasonable agreement between the reconstructed and measured stray fields (Figures 4C-D). For the dipole configuration, the root-mean-square error (RMSE) between the measured and reconstructed stray field profiles was $\SI{1.57}{\milli\tesla}$, and the reconstructed in-plane magnetization averaged $\SI{329.29}{\kilo\ampere/\meter}$ ($\sigma=\SI{261.65}{\kilo\ampere/\meter}$). The Landau domain reconstruction was better, with a stray field RMSE over the line cut of $\SI{0.77}{\milli\tesla}$ and an average magnetization of $\SI{307.79}{\kilo\ampere/\meter}$ ($\sigma=\SI{263.66}{\kilo\ampere/\meter}$). It should be noted that the line profile for the RMSE error, provided in Figures 4E and 4F, are different from the RMSE error from the 2D map.

To illustrate the fundamental challenge of the ill-posed inverse problem, we note that initializing the network without magnetization constraints leads to rapid convergence in $B$-field space while producing physically meaningless magnetization distributions. Without spatial masking or domain initialization, the network exploits the non-uniqueness of the inverse problem, generating arbitrary magnetization patterns that satisfy the forward model but bear no resemblance to physical domain structures such as the four-domain Landau pattern. This behavior underscores that physics-informed constraints are not merely beneficial for model performance but essential for meaningful reconstruction—the network cannot discover physical domain structures from stray field data alone. While the dependence of output magnetizations on user-imposed constraints highlights limitations on quantitative reconstructions, this deficit can be exploited to compare the relative likelihood that a given initial magnetization is correct from the quality of the model's convergence. This work demonstrates that in an experimental setting, such a technique successfully discriminates against predicted magnetizations that do not physically align with the sample.

The NV-to-sample standoff distance, which appears in an exponential term ($e^{-kz}$), is a critical source of uncertainty.  Small errors in estimating this distance, especially in the presence of sample topography, can lead to significant scaling errors in the reconstructed magnetization. The influence of this standoff distance is non-monotonic, creating a complex relationship between experimental uncertainty and reconstruction accuracy. Considering the dipole domain, decreasing the standoff distance can improve the relative error, as observed in simulations where error dropped from $\SI{0.114}{\micro\tesla^2}$ to $\SI{0.102}{\micro\tesla^2}$ when decreasing the distance from 50 nm to 15 nm. This occurs because the exponential decay term $e^{-kz}$ in Fourier space acts as a low-pass filter, preferentially attenuating the high-frequency magnetic features that are most challenging for the network to reconstruct. As the distance decreases further, this filtering benefit is quickly overshadowed by severe signal attenuation. 

For the more complex Landau domain, attenuation of low frequencies hinders the model's ability to recover a solution. Once the signal strength approaches the intrinsic noise floor of the system, the signal-to-noise ratio collapses, causing the reconstruction error to degrade rapidly from $\SI{0.324}{\micro\tesla^2}$ at 50 nm to $\SI{0.354}{\micro\tesla^2}$ at 15 nm. This delicate balance illustrates that an optimal, rather than minimal, standoff distance may exist for achieving the highest fidelity reconstruction. Furthermore, the choice of a spatial filter cutoff frequency for pre-processing the data involves a trade-off between resolving fine features and introducing artifacts, which can also influence the final quantitative values. A detailed analysis is required to fully de-convolve these experimental factors from the reconstruction algorithm's performance.

Finally, while our deep image prior (DIP) approach demonstrates reconstruction of these prototypical domains in Permalloy nanostructures, it is important to recognize the boundaries of its applicability. The method’s reliance on accurate spatial masking is well-suited for systems where domain walls are well understood \textit{a priori} using techniques like MFM or micromagnetic simulations. However, for magnetic textures with more complex or topologically non-trivial features—such as skyrmions or antiferromagnetic domains—the ambiguity in stray field signatures can limit reconstruction fidelity. In these cases, more sophisticated architectures, such as U-Nets with skip connections, have shown success~\cite{Broadway2025, Dubois2022}. These methods are often less dependent on precise prior knowledge of domain boundaries. Nevertheless, our DIP framework offers distinct advantages in computational efficiency and ease of implementation for the routine analysis of technologically relevant materials with well-characterized domain patterns. Future work may integrate elements from both methodologies to further expand the scope of neural network-based magnetic reconstruction.

In summary, we have presented a deep image prior framework for reconstructing in-plane magnetization patterns from scanning NV magnetometry data. Using a simple encoder-decoder architecture, we achieve qualitative reconstructions of Landau and dipole domain structures in Permalloy nanostructures with reasonable quantitative estimations of material parameters. Notably, we find that the sensitivity of the reconstruction to mask orientation can be exploited to gain insight into unknown domain structures. Our results consolidate the understanding that physics-informed neural networks can provide accurate and efficient solutions to the ill-posed inverse problem of magnetization reconstruction, particularly for systems with well-defined domain boundaries. While the present approach is well-suited for prototypical domain patterns such as dipole and Landau domains, further work will be necessary to extend its applicability to more complex or topologically nontrivial magnetic textures. Overall, this framework offers a practical tool for quantitative analysis of magnetic nanostructures and may facilitate improved characterization in device-relevant materials.

\section{Associated Content}
Our NV-NET code is available on GitHub: https://github.com/zsscholl/NVNet

\section{Author Information}
\textbf{Corresponding Author}
\newline Hanu Arava---Materials Science Division, Argonne National Laboratory, IL 60439, USA;https://orcid.org/0000-0003-4866-7616;Email:harava@anl.gov

\textbf{Authors}
\newline Zander Scholl---Materials Science Division, Argonne National Laboratory, IL 60439, USA;ORCID;Email:zsscholl@gmail.com
\newline Justin Woods---Materials Science Division, Argonne National Laboratory, IL 60439, USA;ORCID;Email:jwoods@anl.gov
\newline Charudatta Phatak---Materials Science Division, Argonne National Laboratory, IL 60439, USA;ORCID;Email:cd@anl.gov

\textbf{Author Contributions}
\newline  
\newline \textbf{Notes} The authors declare no competing financial interest.

\begin{acknowledgments}
Z.S. was supported in part by the U.S. Department of Energy, Office of Science, Office of Workforce Development for Teachers and Scientists (WDTS) under the Science Undergraduate Laboratory Internships Program (SULI). H.A, J.W., and C.P. were funded by the US Department of Energy, Office of Science, Office of Basic Energy Sciences, Materials Science and Engineering Division. Work performed at the Center for Nanoscale Materials, a U.S. Department of Energy Office of Science User Facility, was supported by the U.S.DOE, Office of Basic Energy Sciences, under Contract No.DE-AC02-06CH11357. 
\end{acknowledgments}

\bibliographystyle{IllingAJPhack}
\bibliography{references}
\end{document}